\documentstyle[12pt,epsfig]{article}

\def\Title#1{\begin{center} {\Large #1 } \end{center}}

\begin{document}

\begin{flushright}
\begin{displaymath}
\hspace*{11cm}
\begin{array}{l}
$CERN-TH/99-414$\\
$hep-ph/0001134$\\
$December 1999$
\nonumber
\end{array}
\end{displaymath}
\end{flushright}

\vspace*{0.5cm}

\Title{\bf Applications of QCD\footnote{Talk given at 
XIXth International Symposium on Lepton and 
Photon Interactions at High Energies (LP 99), Stanford, California, 9-14 
August 1999.} 
}

\bigskip\bigskip


\begin{raggedright}  

{\it Martin Beneke\index{Beneke, M.}\\
Theory Division, CERN\\
CH-1211 Geneva 23, Switzerland}
\bigskip\bigskip
\end{raggedright}

\section*{Introduction}

Quantum chromodynamics (QCD), the theory of the strong interaction, 
attracts a large body  
of theoretical work and experimental investigation. 
``Applications of QCD'' [theory] probably means that the former should 
have something to do with the latter. This is a severe restriction. 
It leaves out, for example, phenomena at high temperature and large 
baryon density, which have not been tested in terrestrial laboratories. Yet 
there have been fascinating developments in these directions, which 
reflect once more the complexity that follows from a Lagrangian as 
simple as $-G^2/4\,+\,$quarks. In this talk I concentrate on high energy 
QCD processes, which means that at least some part of the process 
must be tractable 
perturbatively. Even within this narrow frame striving at completeness 
would do injustice to the diversity of the (sub-)field. The following 
gives a survey and assessment of recent theoretical results on selected 
topics. For details please consult original references and topical 
reviews. Apologies for omitting topics that should have been 
included, but have not been for various reasons (lack of time, competence, 
...).

The conceptual basis for discussing QCD processes at 
large momentum transfer $Q$ is provided by factorization:
\begin{equation}
\label{ff}
d\sigma = d\hat{\sigma}(Q,\mu) \otimes F(\mu,\Lambda_{\rm QCD}) + 
O(\Lambda_{\rm QCD}/Q).
\end{equation}
The first factor, $d\hat{\sigma}(Q,\mu)$, is insensitive to long distances 
of order $1/\Lambda_{\rm QCD}$. It is computed in 
perturbation theory as scattering of quarks and gluons and depends 
only on the strong coupling $\alpha_s$ and heavy quark masses. The 
second factor accounts for the fact that experiments are prepared 
and measurements are done far away ($\gg 1\,$fm) from the interaction 
point. $F(\mu,\Lambda_{\rm QCD})$ parametrizes this 
long-distance sensitivity in terms of process dependent quantities: 
vacuum condensates, parton distributions, fragmentation functions, 
light cone wave-functions and many more. In principle, 
$F(\mu,\Lambda_{\rm QCD})$ depends only on $\alpha_s$ and light 
quark masses, but since $\alpha_s(\Lambda_{\rm QCD})$ is large, 
we cannot compute it in perturbation theory. However, being
independent of the hard scattering process, the same 
$F(\mu,\Lambda_{\rm QCD})$ may appear in a generic class of
processes. There are some fortunate 
cases in which long distance sensitivity appears only in 
$O(\Lambda/Q)$ in (\ref{ff}). In these cases, we have particularly 
clean predictions, if $Q$ is large enough. In general, we need to 
provide $F(\mu,\Lambda_{\rm QCD})$. It can sometimes be computed 
non-perturbatively by numerical methods (``lattice QCD''). Or it may be 
approximated by models of low-energy QCD. More often, however, 
some measurements are used 
to determine $F(\mu,\Lambda_{\rm QCD})$; others are then predicted. 
This makes QCD seem to depend on many infrared parameters along 
with $\alpha_s$. It also implies iterations of theory and experiment 
to arrive at predictions.

Eq.~(\ref{ff}) suggests a procedure: for any given large momentum 
transfer process (i) establish (\ref{ff}), identify 
$F(\mu,\Lambda_{\rm QCD})$; (ii) compute $d\hat{\sigma}(Q,\mu)$ 
accurately; (iii) if $F(\mu,\Lambda_{\rm QCD})$ is known, predict 
$d\sigma$, otherwise determine $F(\mu,\Lambda_{\rm QCD})$, if 
$d\sigma$ is measured; (iv) check the accuracy of this procedure by 
addressing  power corrections $O(\Lambda_{\rm QCD}/Q)$. The outline 
of this talk is divided in sections according to 
this procedure rather than by 
topics, although in different order. Sect.~\ref{sect1} covers 
perturbative calculations, Sect.~\ref{sect2} power corrections. 
Perturbative expansions of $d\hat{\sigma}(Q,\mu)$ often fail 
in special kinematic regions, but accurate results can be recovered 
upon all-order resummations. In Sect.~\ref{sect3} I discuss three 
representative examples of this situation. Finally, Sect.~\ref{sect4} is 
devoted to some processes for which factorization has been established 
more recently.

It is important to remind ourselves that working with 
QCD we take many things for granted which have never been proven: 
that (\ref{ff}), obtained from factorization properties of Feynman 
diagrams, holds non-perturbatively; that the operator product expansion 
holds non-perturbatively; that perturbative expansions are asymptotic; 
that lattice QCD approaches the correct continuum limit. The overall 
picture of consistency that has emerged in applications of QCD suggests 
a pragmatic attitude towards these problems. However, the questions 
remain.

\section{Perturbative calculations}
\label{sect1}

For long-established QCD processes there are no easy perturbative 
calculations any more. Increasing the accuracy by one order in 
$\alpha_s$ has become technically demanding, usually requiring 
extensive or automated algebraic manipulations by computers and/or 
numerical computing. The complications increase by increasing 
the number of loops, or the number of mass scales or external legs.  

\subsection{More loops}

Totally inclusive quantities are related to imaginary parts of 
correlation functions. This avoids infrared divergences in intermediate 
expressions. Such quantities are candidates for fully automated 
evaluation \cite{Harlander:1999dq}. 

The $\alpha_s^3$ correction to $e^+ e^-\to$ hadrons (massless quarks and 
gluons) and related observables, and to some deep inelastic scattering 
sum rules, have been known for some time \cite{Gorishnii:1991vf}. 
More recently, 
the QCD $\beta$-function \cite{vanRitbergen:1997va} and quark mass 
anomalous dimension 
\cite{Chetyrkin:1997dh} have been computed at 4-loop order.

These results use that any 3-loop, massless, 2-point integral 
is calculable in dimensional regularization. The most important tools 
are the integration-by-parts method \cite{Tka81}, infrared 
re-arrangement \cite{Vla80} which reduces the calculation of the 4-loop pole 
part to the above class of diagrams, and powerful computers 
that handle the algebra connected with about $10^4$ Feynman 
diagrams. Another important class of diagrams which is generically 
calculable is 3-loop, massive, vacuum bubble diagrams 
\cite{Broadhurst:1992fi}. There is 
no obvious way to extend these results to one more loop.

\subsection{More scales}

Observables that depend on more than one kinematic invariant or a 
kinematic invariant and quark masses are difficult, even if they are totally 
inclusive. A method that has led to a number of interesting new results 
is based on asymptotic expansions in a ratio of scales, such that 
each term in the expansion is a single-scale integral that is 
analytically solvable. This method can be used even if the expansion 
parameter is not small if many terms in the expansion can be obtained and 
if the radius of convergence is sufficiently large or convergence can 
be improved by Pad\'{e} approximants.

Asymptotic expansions can be performed (i) for large external momenta, 
small masses or for large masses, small external momenta 
\cite{Chetyrkin:1982zq}; (ii) around 
mass shell \cite{Smirnov:1997ng}; 
(iii) near thresholds \cite{Beneke:1998zp} or 
in $t/s$ for $2\to 2$ scattering; 
(iv) in Sudakov limits \cite{Smirnov:1997gx}. 
These expansions are done on the 
integrand level. The fact that loop 
momenta cover all scales implies that, in general, extra terms have 
to be added to the Taylor expansion of the integrand.

A nice example to illustrate the method is the 3-loop coefficient 
in the relation between the pole mass and the $\overline{\rm MS}$ mass 
of a heavy quark \cite{Chetyrkin:1999ys}. This requires 
3-loop on-shell integrals, 
which are not known. Instead expand the quark self-energy around external 
momentum $p^2=0$, which reduces the problem to 3-loop vacuum bubbles, 
which are calculable. Then put $p^2/m^2=1$ and use Pad\'{e} approximants. 
Expansion to order $(p^2/m^2)^{14}$ (plus information from the 
opposite limit $p^2\gg m^2$) gives $r_3=3.10\pm 0.06$ for the 
coefficient at order $\alpha_s^3$ ($n_f=4$).
(In retrospect, this turns out to be a ``bad'' example, because the 
3-loop on-shell integrals are in fact exactly calculable 
\cite{Laporta:1996mq,Melnikov:2000qh}. 
The exact number is $r_3=3.0451\ldots$ \cite{Melnikov:2000qh}, 
in nice agreement with the 
previous semi-analytic result.)

Applications of this method up to now concern quantities with 
internal masses or on-shell lines. $e^+ e^-\to b\bar{b}X$ has been 
obtained at order $\alpha_s^2$ for general $q^2$ 
\cite{Chetyrkin:1996ii} and 
in an expansion near threshold \cite{Czarnecki:1998vz}. 
The $\alpha_s^2$ corrections to 
inclusive heavy quark decays has been calculated for $b\to c l\nu$ 
\cite{Czarnecki:1997cf}, 
$t\to b W$ \cite{Czarnecki:1999qc} and $b\to u l\nu$ 
\cite{vanRitbergen:1999gs}. The last result 
is particularly impressive, because the asymptotic expansion is 
obtained algebraically to all orders. It is then resummed to an exact 
result.

\subsection{More legs}

Higher order jet calculations pose a different sort of challenge, 
because the kinematics becomes complicated (as the number of jets 
increases), and because the calculation is done on the amplitude 
level. Infrared singularities cancel in an intricate way, or 
are factorized into parton densities (fragmentation functions) after 
cancellations. Almost certainly the final result is obtained after 
numerical integration.

Relatively recent results include NLO corrections to 
$e^+ e^-\to 4$ jets \cite{Signer:1997bf}, $e^+ e^-\to 3$ jets with
quark mass effects 
\cite{Bernreuther:1997jn}, which provide us with a first, yet
imprecise, evidence of 
scale-dependence of the bottom quark mass \cite{Abreu:1998ey}. 
Partial NLO results exist on $p\bar{p}\to 3$ jets \cite{Kilgore:1997sq}. 
The full result 
is supposed to be completed soon.

\subsection{Towards NNLO jets}

The conceptual and technical frontier is set by NNLO jet calculations, 
the basic process being $2\to 2$ ($pp\to 2$ jets or 1 jet inclusive, 
$pp\to \gamma\gamma X$) or $1\to 3$ ($e^+ e^-\to 3$ jets). NNLO 
calculations provide detailed insight into jet structure and a better 
determination of $\alpha_s$. In $e^+ e^-\to 3$ jets they are important 
to understand the interplay between perturbative and power 
corrections.

There are several components to the NNLO jet project. The amplitudes 
have to be computed, which include 2-loop 4-point diagrams. Amplitudes 
with five and six partons have to be integrated analytically over the 
singular regions of phase space. After cancellation of infrared 
divergences, the remaining phase space integrals have to be evaluated 
numerically efficiently.
There has been progress on many of these components recently. 

Because of the integration over singular regions of phase space even 
tree amplitudes are non-trivial. In $2\to 4$ tree amplitudes one 
encounters a new situation, when two partons become simultaneously 
soft or three partons become 
collinear. The last case (squared and integrated over phase space) 
gives rise to a new class of splitting amplitudes (functions), 
when one parton decays into three collinear partons, which generalize 
the usual splitting functions. All soft, collinear, and mixed, limits 
have now been analyzed \cite{Campbell:1998hg}. Likewise, 
although the 1-loop five-point 
amplitudes are known in four dimensions, this is insufficient, because 
the two jet cross sections includes configurations where two of the 
partons are not resolved. Making use of the universality of soft and 
collinear limits, these amplitudes are now known to all orders in the 
dimensional regularization parameter $\epsilon$ in those kinematic regions, 
where the phase space integration is singular \cite{Bern:1998sc}.

The most difficult amplitude is the 2-loop virtual correction to the 
basic $2\to 2$ (or $1\to 3$) process. Until very recently, it has been 
unclear whether the basic scalar double box integrals are analytically 
calculable. In a stunning calculation \cite{Smirnov:1999gc} 
an analytic result 
was obtained for the planar double box integral, expressed in terms of 
elementary special functions, and an algorithm was provided 
to compute the integral
with arbitrary numerator \cite{Smirnov:1999wz}. 
It is equally surprising that this 
result was obtained by elementary methods: the $\alpha$-representation 
and Mellin-Barnes transformation and summations of multiple sums obtained 
after taking the Mellin-Barnes integrations. The crossed double 
box was subsequently calculated \cite{Tausk:1999vh} using the same
methods. The 
numerator algebra, connected to multiple products of three-gluon vertices, 
is, however, highly non-trivial, and remains to be done. Methods, based 
on helicity amplitudes, colour decomposition, special gauges and 
unitarity exist to simplify the task \cite{Bern:1996je}. 
Up to now this has been 
completed in a toy $N=4$ supersymmetric theory (leaving the scalar 
integral unevaluated) \cite{Bern:1997nh}, and more recently for the 
maximally helicity violating amplitude in QCD \cite{Bern:2000dn}.

Many of these results can be used also for NNLO corrections to 
$e^+ e^-\to 3$ jets. However, the 2-loop double box integrals with one 
off-shell external leg are not yet known. The infrared singularities 
at order $1/\epsilon^{4,3,2}$ are known \cite{Catani:1998bh}, 
but the structure of $1/\epsilon$ poles remains to be elucidated.

In my opinion, the results that have been achieved over the past two years 
make success predictable, at least for NNLO 2 jets. On the other hand, 
many hard algebraic and numerical tasks remain to be done. 
Even with a concerted 
effort the relevant time scale is years rather than months. However, this 
is clearly a beautiful case, where most of the most advanced techniques 
for perturbative QCD calculations merge into a single project.

\subsection{NNLO parton evolution}

NNLO jets require NNLO parton distributions. Evolution of these parton 
distributions requires the NNLO DGLAP splitting functions. The complete 
NNLO splitting functions are still unknown, although some moments have 
been computed \cite{Larin:1994vu} some time ago and further
constraints exist in 
the large-$x$ and small-$x$ limit.

Large evolution means large $Q^2$, since parton distributions are typically 
determined experimentally at some low scale. 
Large $Q^2$ means large $x$ and hence the known moments may already 
provide accurate information.
Indeed, first constructions of approximate NNLO non-singlet 
splitting functions have 
appeared that make use of the known information
\cite{vanNeerven:1999ca}. The constraints at 
small $x$ turn out to be quite weak, but there seems to be little 
uncertainty for $x>0.1$. When the splitting function is folded with a 
typical parton distribution this range increases to $x>0.02$.

\section{Beyond leading power}
\label{sect2}

Perturbative expansions, if computed to arbitrarily high order, 
ultimately diverge. They become useless beyond a certain order, unless they 
are summed. In recent years, we have learnt to turn this embarrassment 
into a benefit, since the pattern of divergence tells us something 
about the scaling of power corrections $(\Lambda_{\rm QCD}/Q)^n$ to 
a hard scattering cross section. A particularly interesting type 
of divergence, called infrared renormalon, is related to integration 
over small loop momentum in Feynman integrals \cite{thooft}. 
Roughly speaking, there is a relation 
between perturbative long distance sensitivity, the size 
of perturbative coefficients in higher orders, and the scaling of 
non-perturbative power corrections \cite{Beneke:1999ui}. 
For inclusive deep inelastic 
scattering, $n=2$, and one recovers higher twist corrections 
predicted by the operator product expansion. 

\subsection{Event shape observables and energy flow}

For other, less inclusive, observables, such as event shape variables 
in $e^+ e^-$ and $ep$ collisions, one often finds $n=1$ 
\cite{Manohar:1995kq,Dokshitzer:1995zt,Nason:1995hd}. Since these  
variables are order $\alpha_s$ perturbatively, they are prone to 
large non-perturbative (and perturbative) corrections. They have been 
investigated intensively over the past two years, theoretically and 
experimentally. 

The leading power correction originates from soft partons emitted from 
a fast, nearly back-to-back $q\bar{q}$ pair. Write 
\begin{equation}
\langle S\rangle = \langle S^{\rm pert.}(\mu_I)\rangle + 
\frac{\mu_I}{Q}\,\langle S^{NP}(\mu_I)\rangle + 
O(\Lambda_{\rm QCD}^2/Q^2)
\end{equation}
for an average event shape variable $S$. The experimentally measured 
energy dependence of $\langle S\rangle$ clearly supports the existence 
of a $1/Q$ power correction with a reasonably sized normalization 
$\langle S^{NP}(\mu_I)\rangle$, which is non-perturbative. An interesting 
hypothesis (also applied to event shape distributions 
\cite{Dokshitzer:1997ew}) 
states that the non-perturbative corrections are universal, 
i.e. $\langle S^{NP}(\mu_I)\rangle\propto c_S \overline{\alpha}(\mu_I)$, 
where $c_S$ is observable dependent, but calculable and 
$\overline{\alpha}(\mu_I)$ is non-perturbative but independent of $S$ 
\cite{Dokshitzer:1995zt}. (This applies to thrust, jet masses and the
$C$ parameter. Other 
event shapes, such as jet broadenings, involve complications 
\cite{Dokshitzer:1999qp}.) 
Four parton final states with two soft partons have also been 
investigated \cite{Dokshitzer:1998iz,Beneke:1997sr,Dasgupta:1999mb}. 
Remarkably, one finds that $c_S$ is rescaled by 
the same factor for a variety of shape observables
\cite{Dokshitzer:1998iz,Dasgupta:1999mb}. 
The universality 
hypothesis has led to a number of instructive experimental tests. 
Recent results on average event shapes and event shape distributions 
in $e^+ e^-$ annihilation 
\cite{MovillaFernandez:1999yn} and DIS \cite{Rabbertz:1999hp} 
tend to confirm the hypothesis within 
the expected accuracy. However, the fact that the value of $\alpha_s$, 
fitted simultaneously to each $S$, is somewhat unstable indicates that 
the present understanding is not perfect.

Universality may hold for a special class of observables, but it would be 
surprising, if it held in general. What is needed to shed light on the 
issue is a factorization theorem for soft gluons beyond leading 
power. Recall that factorization theorems for event shapes usually 
demonstrate that soft gluon corrections cancel at leading power. We 
are now interested in the leading contribution that is left over 
after this cancellation. 

In \cite{Korchemsky:1997sy} the problem is approached in terms 
of energy flow of soft 
particles. The universal, non-perturbative objects relevant to the 
two-jet limit ($q\bar{q}$ plus soft partons) are 
\begin{equation}
G(\vec{n}_1,\ldots,\vec{n}_k;\mu_I) = \langle 0|W^\dagger 
\prod_{i=1}^k {\cal E}(\vec{n}_i) \,W|0\rangle,
\end{equation}
where ${\cal E}(\vec{n}_i)$ measures soft energy flow in the direction 
of $\vec{n}_i$, $\mu_I$ is a factorization scale that defines what 
``soft'' means, and $W$ denotes a product of eikonal lines for the 
energetic $q\bar{q}$ pair. The $G(\vec{n}_1,\ldots,\vec{n}_k;\mu_I)$ are 
horribly complicated objects and it is hardly 
conceivable that they could ever be extracted from measurements. 
However, the fact that they are independent of the hard scale $Q$ already 
entails interesting predictions. For example, event 
shape distributions can be expressed as a convolution of a perturbative 
distribution and a non-perturbative 
$Q$-independent, but observable-dependent ``shape 
function'', that follows from these energy flow correlation functions. 
Event shape averages can be represented as 
\begin{equation}
\langle S\rangle_{1/Q} = \int d\vec{n}\,w_S(\vec{n})\,G(\vec{n}),
\end{equation}
with a calculable weight function $w_S(\vec{n})$. The single energy flow 
correlation function $G(\vec{n})$ can in principle be determined 
from the leading power 
correction to the energy-energy correlation. 

I find this a promising step towards understanding soft power corrections. 
The concept of energy flow is clearly important and deserves more attention, 
as it corresponds directly to calorimetric measurements. Observables that 
can be represented in terms of energy flow are automatically infrared 
safe. They may also be defined non-perturbatively and therefore be 
amenable to a more systematic analysis of power corrections 
\cite{Sveshnikov:1996vi}.

\subsection{OPE of the plaquette}

There are also things that don't work as expected. Consider the 
operator product expansion (OPE) of the plaquette expectation value 
in pure gauge theory at finite lattice spacing, i.e. the inverse 
lattice spacing takes the role of the scale $Q\gg \Lambda_{\rm QCD}$. 
The OPE gives
\begin{equation}
\langle \mbox{plaquette}\rangle = 
\sum_{n=1} c_n\alpha^{\rm latt}_s(Q)^n + \frac{C(\alpha^{\rm latt}_s)}
{Q^4}\,\langle \frac{\alpha_s}{\pi} GG\rangle + \ldots.
\end{equation}
The coefficients $c_n$ of the perturbative expansion have been computed 
to 8th order numerically \cite{DiRenzo:1995qc}. After transformation 
to a continuum coupling 
definition, the coefficients exhibit the expected infrared renormalon 
growth. Summing the series approximately should give an accuracy of 
order $1/Q^4$, hence subtracting the summed series from non-perturbative 
Monte Carlo data for $\langle \mbox{plaquette}\rangle$, the remainder 
should scale as $1/Q^4$, consistent with the scaling of the gluon 
condensate term.

Contrary to this expectation, the remainder is found to approach 
a perfect $1/Q^2$ scaling behaviour \cite{Burgio:1998hc}. 
Since the OPE is one of 
the few tools we have to go beyond perturbation theory, this is clearly 
something we should understand. There may be subtleties with the 
transformation to the continuum scheme, since this transformation 
is not known to 8th order or may also have power corrections. The 
effective action at finite lattice spacing contains an infinite set of 
higher dimension operators. Could these add up to a $1/Q^2$ power 
correction \cite{Beneke:1998eq} so that the result is a lattice
artefact? But there 
may be less profane explanations such as power corrections from short 
distances that affect coefficient functions (and therefore would not 
contradict the OPE) \cite{Akhoury:1998by}. 
This possibility is not ruled out by any 
argument. It presents a fundamental question that challenges our 
understanding of non-perturbative short-distance expansions. It would 
also have implications for the phenomenology of power corrections 
to current correlation functions. For these reasons, the problem 
raised by \cite{Burgio:1998hc} should be cleared up!

\section{Perturbative resummations}
\label{sect3}

Returning to perturbative expansions, it is not unusual that a 
perturbative expansion in $\alpha_s$ breaks down, even though 
the coupling constant is small. This happens because the smallness 
of the coupling constant is compensated by a large kinematic 
invariant. In effect, one is dealing with a multi-scale problem. 
If all scales are large compared to $\Lambda_{\rm QCD}$, the problem 
is perturbative and may be subjected to systematic all-order 
resummations. The kinematic conditions leading to the breakdown 
of perturbation theory can be quite different and the resummations 
reflect completely different physics. In this section I discuss three 
examples of such resummations, where progress has been made over the 
past two years.

\subsection{Parton thresholds}

A familiar source of large kinematic corrections is related to 
partonic thresholds. Consider the differential cross section 
\begin{equation}
d\sigma = \sum_{i,j} f_{i/A}\otimes f_{j/B}\otimes d\hat{\sigma}_
{ij\to f}
\end{equation}
for a hard hadron-hadron collision. Large logarithms appear in 
$d\hat{\sigma}$, when the cms energy $\hat{s}$ of $i+j$ is just 
large enough to produce a given final state. For example, in 
production of a massive vector boson with mass $Q$, the leading 
correction is $\alpha_s^n (\ln^{2 n-1}(1-z))/[1-z]_+$ at order 
$\alpha_s^n$, where 
$z=Q^2/\hat{s}$, and perturbation theory breaks down for $z\to 1$. 

In this case large logarithms originate from the lack of phase 
space for real emission and the incomplete cancellation of 
sensitivity to collinear and soft momentum. Because of this relation 
the structure of these logarithms is well understood. The logarithms 
exponentiate and can be resummed:
\begin{eqnarray}
\label{expnll}
\int d z\,z^{N-1}\,d\hat{\sigma}(z) &=&
\nonumber\\ 
&&\hspace*{-3.5cm}
H(\alpha_s) \,\exp \left[\ln N\,g_1(\alpha_s\ln N) + g_2(\alpha_s\ln N) + 
\alpha_s g_3(\alpha_s \ln N) + \ldots\right] + O(1/N).
\end{eqnarray}
This resummation was worked out at next-to-leading logarithmic 
order (i.e. 
including $g_2(\alpha_s \ln N)$) some years ago for $2\to 1$ 
processes (massive vector boson production) \cite{Sterman:1987aj} and 
$1\to 2$ processes (event shape variables in $e^+ e^-$ in the 
2-jet limit) \cite{Catani:1993ua}.

Next-to-leading logarithmic resummation has now been extended 
to $2\to 2$ scattering processes 
\cite{Kidonakis:1996aq,Bonciani:1998vc}. 
Several new complications 
appear in this case. Since the underlying hard process depends on 
an additional kinematic invariant, $(-\hat{t})/\hat{s}$, so do 
the functions that appear in the exponent of (\ref{expnll}). 
Furthermore, the $2\to 2$ amplitude contains several colour 
amplitudes and since soft gluon emission carries away colour, these 
amplitudes mix, turning the exponential into a matrix exponential 
on the independent colour amplitudes. While the structure of 
resummation remains thus the same, the technical complications 
make the formalism more difficult to apply in practice.

Fortunately, simplifications occur for total cross sections. NLL resummed 
results have been presented for heavy quark production
\cite{Bonciani:1998vc,Kidonakis:1999ei} and 
prompt photon production \cite{Catani:1999hs}. 
For di-jet production at large 
transverse momentum, the formalism is in principle complete, but 
it has not yet been implemented \cite{Kidonakis:1998bk}. 
It turns out that at energies of 
interest for heavy quark production and prompt photons, the effect of 
resummation is typically small, i.e. within the renormalization scale 
variation of a fixed order NLO calculation. The real benefit of resummation 
is a significant reduction of this scale dependence compared to NLO QCD, 
and hence, probably, the theoretical uncertainty. The $E_T$ spectrum 
of prompt photons at low $E_T$ remains in disagreement with the data 
\cite{Apanasevich:1998hm}. 
Since at $E_T\approx$ several GeV power corrections in $1/E_T$, or 
intrinsic transverse momentum, can be very important, this is hardly 
a serious issue. It is, however, a serious problem for determining 
the gluon distribution at large $x$.

\subsection{Non-relativistic}

A different kind of partonic threshold is encountered in heavy 
quark production in $e^+ e^-$ annihilation. When the cms energy is just 
larger than $4 m_Q^2$, the quark and antiquark move at small relative 
velocity and attract each other through a strong Coulomb force, 
even if $\alpha_s$ is small. Formulated as a perturbative resummation 
problem, we need the terms 
\begin{equation}
R_{e^+ e^-\to Q\bar{Q} X} \sim v \sum_{n=0}^\infty \left(\frac{\alpha_s}
{v}\right)^n \cdot \left\{1 (\mbox{LO}); \alpha_s, v (\mbox{NLO}); 
\alpha_s^2, \alpha_s v, v^2 (\mbox{NNLO});\ldots\right\}
\end{equation}
at leading order (LO), next-to-leading order (NLO), etc., where $v$ is 
the small relative velocity. 

The LO resummation is done by solving for the Green function 
of the Schr\"odinger equation with the Coulomb potential. To be more 
systematic, such concepts from quantum mechanics have to be derived from 
QCD, incorporating correctly the short-distance structure of QCD. This 
is done by a sequence of non-relativistic effective field theories. 
Quarks and gluons can be classified as hard (h), soft (s), potential (p) 
and ultrasoft (us) \cite{Beneke:1998zp}. Then these modes are integrated out 
successively, according to the scheme 
${\cal L}_{\rm QCD}[Q(h,s,p);g(h,s,p,us)] \to 
{\cal L}_{\rm NRQCD}[Q(s,p);g(s,p,us)] \to 
{\cal L}_{\rm PNRQCD}[Q(p);g(us)]$, passing from QCD to 
non-relativistic QCD \cite{Thacker:1991bm} to
potential-non-relativistic 
QCD \cite{Pineda:1998bj}.
The equation of motion of PNRQCD is exactly the Schr\"o\-dinger equation, 
with corrections to it that encode the information of the short-distance 
modes that have been integrated out.

With the help of this method the NNLO resummation has been performed. 
This leads to first principle NNLO 
calculations of $t\bar{t}$ production near 
threshold (in $e^+ e^-$ collisions) 
\cite{Hoang:1998xf,Beneke:1999qg,Hoang:1999zc,Nagano:1999nw}. 
The NNLO correction 
turns out to be very important and has led to the conclusion 
that it is the $\overline{\rm MS}$ top quark mass rather than the pole 
mass than can be determined more accurately, though 
indirectly, from the cross section 
near threshold \cite{Beneke:1999qg,Hoang:1999zc}.  
Another important application concerns the 
determination of the $b$ quark mass from $e^+ e^-\to b\bar{b} X$ 
\cite{Penin:1998zh,Melnikov:1999ug}. The recent analyses 
\cite{Melnikov:1999ug} 
that take care of adequate bottom mass 
renormalization prescriptions converge towards a common value 
for the bottom quark $\overline{\rm MS}$ mass, which I average 
as $\overline{m}_b(\overline{m}_b)=4.23\pm 0.08\,$GeV. 
The centre of attention is now on understanding 
logarithmic corrections in $v$ \cite{Beneke:1999qg,Brambilla:1999xj}.

\subsection{High energy, small $x$}

The high energy limit $s\gg Q^2$ of QCD cross sections is an old, 
yet unsolved problem. Large logarithms can appear either in the 
high-energy limit of hard partonic reactions, such as in 
$\gamma^\star\gamma^\star$ scattering or forward jet production, 
or in the small-$x$ behaviour of parton distributions and their 
evolution. The leading logarithms $(\alpha_s\ln s/Q^2)^n$ have been 
summed long ago by means of the BFKL equation \cite{Kuraev:1977fs}. 
This leads 
to cross sections that rise as $s^{\bar{\alpha}_s \,4 \ln 2}$ ($\bar{\alpha_s}
=N_c\alpha_s/\pi$) with energy. For many years most theoretical 
work has been concerned with the physical mechanism that would make 
the high energy limit compatible with unitarity, but a quantitative 
theory has not yet emerged. Most of the recent 
activity in small-$x$ physics has however been inspired by the completion 
of the NLO correction to the BFKL kernel \cite{Fadin:1998py}, and its 
interpretation. The following discussion concentrates on 
this aspect.

Recall that phenomenological applications of LO BFKL theory have remained 
ambiguous or unsuccessful. HERA data on the gluon density indicates that 
DGLAP evolution works well, in fact too well, down to $x\sim 10^{-6}$. 
No resummation of $\ln x$ corrections to the evolution kernels is 
required. There is some flexibility in the input gluon distribution, 
nevertheless the message is that departures from DGLAP cannot be large. 
Virtual photon scattering has been measured at LEP
\cite{Acciarri:1999ix}. Even allowing 
for the fact that LO BFKL may not predict the normalization of the 
cross section well, the observed energy dependence is less steep 
than predicted. Forward pion production at HERA \cite{Adloff:1999zx} 
may be described by LO BFKL, but other interpretations of the data
seem possible.

It is therefore clearly interesting to see how NLO corrections affect 
this comparison. In the high energy limit, the cross section 
factorizes schematically as
\begin{equation}
\label{smallx}
\sigma = \int \frac{d^2 k_1}{k_1^2}\,\Phi_A(k_1)\,
\frac{d^2 k_2}{k_2^2}\,\Phi_B(k_2)\,\int\frac{d\omega}{2\pi i} 
\left(\frac{s}{k_1 k_2}\right)^\omega G_\omega(k_1,k_2),
\end{equation}
where $A$ usually represents a virtual photon and $B$ a virtual photon 
or a proton. In the latter case the impact factor $\Phi_p(k_2)$ is not 
perturbatively calculable. $k_{1,2}$ denote transverse momenta of the 
scattering objects, $k\sim Q$ for virtual photons, $k\sim \Lambda_{\rm QCD}$ 
for protons. The factorized form (\ref{smallx}) is believed to hold 
to next-to-leading logarithmic order, but beyond this order there 
are terms that cannot be associated with the 
four-(reggeized)-gluon Green function 
$G_\omega(k_1,k_2)$. $G_\omega(k_1,k_2)$  satisfies the BFKL 
equation~\cite{Kuraev:1977fs}
\begin{equation}
\omega G_\omega(k_1,k_2) = \delta^{(2)}(k_1-k_2) + \int \frac{d^2k}{\pi} 
K_\omega(k_1,k)\,G_\omega(k,k_2).
\end{equation}
Roughly speaking, the leading order kernel $K_\omega(k_1,k)$ sums 
a single gluon ladder exchanged between $A$ and $B$ with emissions 
ordered in longitudinal momentum. The NLO correction has to account for 
all configurations in which one power of $\ln x$ is lost. Partial 
results have been collected over many years and the full NLO correction 
has finally been completed \cite{Fadin:1998py}. 
It is usually presented through the action of the kernel on a set of 
test functions:
\begin{equation}
\label{act}
\int d^2 k'\,K_\omega(k,k')\left(\frac{{k'}^2}{k^2}\right)^{\gamma-1}
= \bar{\alpha}_s\chi_0(\gamma) \left[1-\beta_0\bar{\alpha}_s
\ln\frac{k^2}{\mu^2}\right] + \bar{\alpha}_s^2 \chi_1(\gamma).
\end{equation}
In the saddle point approximation for the inverse Mellin integrals, 
treating $\bar{\alpha}_s$ as small, the energy growth $s^\lambda$  
of hard high energy cross sections is then determined by
\begin{equation}
\lambda = \bar{\alpha}_s \chi_0(1/2)+\bar{\alpha}_s^2\chi_1(1/2) = 
\bar{\alpha}_s\,4\ln 2 \left[1-6.5\bar{\alpha}_s\right],
\end{equation}
ignoring the scale dependent part of the kernel. The NLO correction is huge, 
large enough to modify qualitatively the conclusions drawn from leading 
order, which is good. At the same time, the NLO kernel taken at 
face value leads to non-sense results \cite{Ross:1998xw}, unless 
$\bar{\alpha}_s\leq 0.05$, which is unrealistically small.

Much effort has gone into the question whether the NLO result 
invalidates the BFKL resummation programme as a whole. To answer 
this question one has to go beyond a systematic resummation of 
high energy logarithms. Such a step is unavoidably ambiguous and needs to be 
motivated by physics arguments. It appears that much of the 
NLO characteristic function $\chi_1(\gamma)$, even near $\gamma=1/2$, 
can be understood from the singularities at $\gamma=0$ and $1$. 
Note from (\ref{act}) that these singularities correspond to transverse 
logarithms. The leading singularities $1/\gamma^{3,2}$, 
$1/(1-\gamma)^{3,2}$ are related to the symmetric energy scale 
$k_1k_2$ chosen in (\ref{smallx}), the running coupling and the 
non-singular terms of the LO DGLAP splitting function. Remarkably, 
$\chi_1(\gamma)$ is extremely well reproduced just by keeping these 
singularities.

This suggests that these singularities should be summed to all 
orders. Unphysical transverse logarithms generated by the symmetric energy 
scale can be removed \cite{Salam:1998tj} by replacing
\begin{equation}
\chi_0(\gamma) \to \chi_0^\omega(\gamma) = 2 \psi(1)-
\psi(\gamma+\omega/2)-\psi(1-\gamma+\omega/2).
\end{equation}
Although not unique, this seems to be a particularly natural choice. 
After performing this replacement, the NLO correction is 
reduced, though 
not small. Further support for this resummation arises from the 
possibility to introduce a ``rapidity veto'' $y_{i+1}-y_i>\Delta$ 
\cite{Andersson:1996ju}, which is essentially 
a hard cut-off on the momentum region, where the ordering in rapidity 
was not a good approximation in the first place. After resummation, 
the $\Delta$-dependence is small and the NLO correction moderate for 
all $\Delta$ \cite{Forshaw:1999xm}, 
which indicates that the resummed kernel is less 
sensitive to momentum regions where the approximations necessary to 
derive it are not valid.

The remaining $\gamma$-singularities are double poles. Two further 
modifications beyond NLO small-$x$ logarithms need to be performed to 
take care of them. 
First, rather than improving the DGLAP anomalous dimension by 
small-$x$ logarithms, we can take the opposite point of view 
and improve $\chi(\gamma)$ by taking into account all information on 
collinear logarithms 
\cite{Ciafaloni:1999iv,Ciafaloni:1999yw,Altarelli:1999vw}. 
In this way one can arrange, in addition, 
for momentum conservation, which requires vanishing anomalous dimension  
at $\omega=1$. Second, the 1-loop evolution of $\alpha_s$ 
can be taken into account exactly, rather than perturbatively as in 
(\ref{act}). There are two cases to consider, symmetric processes with
$k_1\sim k_2\gg\Lambda_{\rm QCD}$ \cite{Kovchegov:1998ae} and
asymmetric processes \cite{Thorne:1999rb,Ciafaloni:1999yw}.
In the latter case, 
with $Q\sim k_1\gg k_2\sim \Lambda_{\rm QCD}$ as for deep inelastic 
scattering, one must also apply collinear factorization to the 
four gluon Green function, such that \cite{Ciafaloni:1999au}
\begin{equation}
G_\omega(k_1,k_2) = F_\omega^{\rm UV}(k_1)\cdot 
F_\omega^{\rm IR}(k_2) + O(k_2^2/k_1^2).
\end{equation}
The dependence on the non-perturbative low momentum evolution of the 
running coupling is factorized into $F_\omega^{\rm IR}(k_2)$, 
which can be absorbed into the input gluon distribution. This part 
remains beyond perturbative control, although it may well control the 
actual small-$x$ behaviour of the gluon distribution. On the other 
hand, only $F_\omega^{\rm UV}(k_1)$ is $Q$-dependent and hence determines 
the evolution of the gluon density. 

There seems yet not to be an unanimous opinion on which of these aspects 
is most important. For example, \cite{Altarelli:1999vw} argues 
that $\lambda$ 
should be considered as a non-perturbative parameter, while 
\cite{Ciafaloni:1999yw}
takes a less agnostic attitude. 
Ref.~\cite{Thorne:1999rb}, on the other hand, emphasizes the role 
of the running coupling, demonstrating that the effective scale of 
$\alpha_s$ in the anomalous dimension increases as $x$ decreases, 
because of ultraviolet diffusion. It is also claimed that this leads to 
an improved fit to structure function data compared to a standard 
DGLAP fit, which is definitely interesting. Despite these  
different viewpoints, theory clearly seems to be on the right track, 
as the results consistently point towards a smaller (but positive) 
hard pomeron 
intercept compared to LO BFKL. The resummed gluon 
anomalous dimension is also close to the DGLAP one down to rather small 
moments. It will be interesting to see consolidation of this field 
and the first true NLO+improved BFKL predictions for physical 
processes (which needs as yet unknown NLO impact factors).

\section{Novel factorization ``theorems''}
\label{sect4}

In the past sections I discussed hard scattering processes which 
have been known as such. But for other processes factorization of its 
short-distance part has been established only recently. Often 
factorization comes at the expense of introducing new non-perturbative 
parameters. Even if these parameters are not accessible immediately, 
much is gained in terms of conceptual clarity. In this section 
I discuss three examples of such ``new'' applications of QCD.

\subsection{Hard diffraction}

A particularly nice example is hard diffraction \cite{talk}. Discovered in 
hadron-hadron collisions by UA8 about a decade ago \cite{Bonino:1988ae}, 
after the inspiring work of \cite{Ingelman:1985ns}, 
the extent to which hard diffraction is a hard process, 
has remained rather unclear. This has changed completely with the 
arrival of accurate data on hard diffraction in $ep$ scattering 
\cite{Derrick:1993xh}, 
the demise of Regge terminology, and the realization that hard 
diffraction in DIS can be described in close analogy with inclusive 
DIS \cite{Trentadue:1994ka,Berera:1996fj}.

In hard diffractive DIS, $\gamma^*p\to X p$, the proton scatters 
(quasi-)elastically off a virtual photon, which fragments into a 
colour neutral cluster $X$. The scattered proton is usually not 
detected, but since it typically loses only a small fraction of its 
momentum, the event is identified by a large gap in rapidity between 
$p$ and $X$. About 10\% of all DIS events are rapidity gap events. 
Furthermore, hard diffraction is not suppressed with $1/Q^2$ 
relative to inclusive DIS. 

In close analogy with inclusive DIS, the diffractive cross section 
factorizes into a short-distance cross section and a diffractive 
parton distribution \cite{Berera:1996fj,Grazzini:1998ih}:
\begin{equation}
\frac{d\sigma^D(x,Q^2,\xi,t)}{d\xi dt} = 
\sum_{i=q,g}\int\limits_x^\xi dy\,\hat{\sigma}^{\gamma^* i}
(Q,x,y;\mu)\,\frac{d f_i^D(y,\xi,t;\mu)}{d\xi dt}.
\end{equation}
The diffractive parton distribution $f_i^D(y,\xi,t;\mu)$ 
represents the probability to find parton $i$ in the proton with 
momentum fraction $y$ under the condition that the proton stays intact 
and loses longitudinal momentum fraction $\xi$. Note that this definition 
makes no reference to Regge factorization or the pomeron. Neither does it 
make reference to a rapidity gap $\Delta y$ , which follows from 
kinematics alone when $\xi$ is small: $\Delta y \sim \ln (1/\xi)$. The hard 
scattering occurs on a single parton as in ordinary DIS. The dynamics 
that is responsible for the formation of a colour-singlet cluster 
is non-perturbative and therefore 
part of the definition of the diffractive parton distribution.

The physical picture of hard diffraction is perhaps clearest in the 
proton rest frame and reminiscent of the ``aligned jet model'' 
\cite{Bjorken:1973gc,Nikolaev:1991ja}. In the 
proton rest frame, at small Bjorken $x$, the virtual photon splits into a 
$q\bar{q}$ pair long before it hits the proton. The $q\bar{q}$ 
wave-function of the virtual photon suppresses configurations in which 
one of the quarks carries almost all momentum. Yet it is these 
configurations that give rise to a large diffractive cross section, 
because the wave-function suppression is compensated by the large cross 
section for the scattering of a $q\bar{q}$ pair of hadronic transverse 
size off the proton. The harder of the two quarks is essentially a 
spectator to diffractive scattering. The scattering of the softer quark 
off the proton is non-perturbative and cannot be described by exchange 
of a finite number of gluons. Hence there is an unsuppressed probability 
that the softer quark leaves the proton intact. 
This explains the leading twist 
nature of hard diffraction. The details of the scattering of the softer 
quark off the proton are encoded in the diffractive quark 
distribution. In a similar way, the $q\bar{q}g$ configuration in the 
virtual photon, in which the $q\bar{q}$ pair carries almost all momentum,  
gives rise to the diffractive gluon distribution.

Because the short-distance cross section $\hat{\sigma}^{\gamma^\star i}$ 
of hard diffractive DIS is identical to inclusive DIS, the evolution 
of the diffractive parton distributions is identical to those of ordinary 
parton distributions. It follows that the characteristics of diffraction 
are entirely contained in the input distributions at a given scale. It 
is therefore interesting to model these distributions. The original idea 
of a partonic content of the pomeron \cite{Ingelman:1985ns} can be 
interpreted as an 
ansatz in which the diffractive parton distribution factorizes into a 
pomeron flux factor, which determines the $\xi$ dependence, and a parton 
distribution in the pomeron which depends only on $\beta=x/\xi$. The 
precise data from HERA do not support this simple ansatz any more, 
although the problem can be fixed by adding more Regge poles. 
More recent approaches model the proton field off which the Fock states 
of the virtual photons scatter. The semi-classical approach 
\cite{Buchmuller:1996mr}, which 
preceded the factorization theorem, can be formulated in such a way 
that it models the diffractive parton distributions
\cite{Hebecker:1997gp}. It can 
be justified for a large nucleus \cite{Hebecker:1998kv}. 
Applied to the proton it gives a 
reasonable description of both diffractive and inclusive DIS 
\cite{Buchmuller:1999jv}. (See \cite{Nachtmann:1991ua} for 
earlier work that contains some elements of the semi-classical 
approach.) Another 
approach is based on two gluon exchange 
\cite{Nikolaev:1991ja,Nikolaev:1994th}. 
In this case one either has 
to deal with an infrared divergence, or couple the gluons to a 
small size toy nucleon as in \cite{Hautmann:1998xn}. Remarkably, 
these three approaches 
give similar results on the $\beta$-dependence of diffractive parton 
distributions and agree on the fact that the gluon distribution is 
enhanced by a large colour factor. This leads to positive scaling 
violations already at relatively large $\beta$, different from inclusive 
DIS, but in agreement with data. 

It is encouraging that simple models reproduce the gross features 
of the data. Given the differences of the models as far as the proton 
is concerned, it seems that hard diffraction probes the wave-function of 
a virtual photon rather than the structure of the proton!

Hard diffraction in hadron-hadron collisions is much harder to describe 
and more varied, as there can be rapidity gaps between jets, between 
a jet and a hadron remnant etc.. Factorization does not seem to hold 
in this case \cite{Alvero:1999ta}, 
neither is it expected to \cite{Berera:1996fj}, since, 
for example, an elastically 
scattered hadron must traverse the remnant of the other hadron, which 
can cause its break-up. I would like to note, however, a recent 
suggestion \cite{Oderda:1998en} to describe rapidity gap-like events 
(between jets) 
in terms of small energy flow in the gap rather than the absence of 
particles. Although this does not correspond exactly to the notion 
of hard diffractive scattering, such a definition is more appropriate 
for a partonic interpretation.

\subsection{Skewed processes}

Factorization has also been shown for deeply virtual Compton scattering 
$\gamma^\star p\to \gamma p$ \cite{Muller:1994fv} 
and diffractive vector meson production \cite{Collins:1997fb} 
(after earlier work 
in \cite{Ryskin:1993ui}) $\gamma(Q) p\to V p$, where $V$ can be an onium and 
$Q$ arbitrary or $V$ can be a longitudinally polarized light vector 
meson, in which case $Q^2$ must be large. Note that two-gluon 
exchange is applicable to diffractive vector meson production, 
but not to diffractive DIS, because convolution with the virtual photon 
wave-function relevant to longitudinal vector meson 
production suppresses the asymmetric $q\bar{q}$ fluctuations, 
which have large transverse size. As a consequence, 
only the small size $q\bar{q}$ component contributes at leading power. 

Deeply virtual Compton scattering and diffractive vector meson 
production require a generalized parton distribution on the amplitude 
level, since the proton is scattered with non-zero momentum transfer, 
owing to the difference in invariant mass of the initial and final 
vector particle. These objects, defined as 
\begin{equation}
p^+\int\frac{d z^-}{2\pi}\,e^{i x p^+ z^-}\,\langle p'|\overline{\psi}
(0)\gamma^+\psi(z^-)|p\rangle
\end{equation}
for quarks,  
are referred to as skewed (off-diagonal, non-forward, ...) parton 
distributions, and describe a parton $i$ (a quark above) extracted from 
the proton with momentum fraction $x$ and returned with momentum 
fraction $x'$. For $p'=p$, the skewed parton distribution reduces to 
the conventional one. The first moment, however, is related to a proton 
form factor. These hybrid properties are also reflected in the evolution 
properties. For $x'>0$ the evolution resembles DGLAP evolution. For 
$x'<0$, the skewed parton distribution describes emission of a 
$q\bar{q}$ pair and the evolution resembles ERBL 
\cite{Efremov:1980qk} evolution of 
light cone distribution amplitudes. The evolution properties  
and the form of skewed parton densities have been actively studied. 
An interesting observation is that the skewed parton density is 
determined by the conventional one for small $x$ and $x'-x$ 
\cite{Shuvaev:1999ce}.

Is there experimental evidence for skewedness? ZEUS \cite{DCVS} reports 
first evidence for deeply virtual Compton scattering, but the data are not 
yet good enough to allow detailed tests. Skewedness effects in 
diffractive vector meson production are largest if the invariant mass 
difference between the incoming photon and outgoing vector meson is 
large. This suggests to look at $\Upsilon$ photoproduction 
\cite{Frankfurt:1999yf} 
or vector meson production at large $Q^2$ \cite{Martin:1999wb}. 
Incorporation of 
skewedness improves the theoretical prediction in comparison with data, 
but other theoretical uncertainties remain large and preclude an 
unambiguous statement. The power behaviour of longitudinal to 
transverse $\rho$ meson production appears to disagree with the 
naive estimate $\sigma_L/\sigma_T\sim Q^2$, but the (formally
logarithmic) scale-dependence 
of the gluon distribution may play an important role 
\cite{Martin:1999wb}.

\subsection{Exclusive $B$ decays}

There exists a standard framework to discuss exclusive processes 
at large momentum transfer in terms of light cone distribution 
amplitudes \cite{Efremov:1980qk}. 
As we are entering the era of exclusive $B$ decays, it is 
only appropriate to consider them as bona fide hard reactions. After 
all they involve momentum transfers 
$q^2\sim m_b^2\sim 25\,\mbox{GeV}^2$, and there will be millions of 
$B$'s! A general two-body  decay amplitude can be written as 
\begin{equation}
\label{amp}
{\cal A}(B\to M_1 M_2) = A_1 e^{i\delta_1} e^{i\delta_{W1}} + 
A_2 e^{i\delta_2} e^{i\delta_{W2}},
\end{equation}
where $A_{1,2}$ denote the magnitudes of the amplitudes, $\delta_{1,2}$ 
strong interaction phases and $\delta_{W1,2}$ weak phases. The 
weak phases are CP-violating and of primary interest. Yet to determine 
them, the strong phases and amplitudes must be known, unless we 
are fortunate enough that only a single term contributes on the 
right hand side of (\ref{amp}), or we have enough experimental information 
to reduce strong interaction input. 

The standard formalism does not immediately apply to 
$B$ decays, because the $B$ meson contains a soft spectator quark. 
The spectator quark may go to a final state meson without 
participating in a hard scattering. Hence the process cannot be described 
in terms of light cone distribution amplitudes alone. A more general 
factorization for decays into a heavy ($D$) and a light meson 
has been proposed in \cite{Bjorken:1989kk}, 
based on the idea that the light meson is 
initially ejected as a compact object from the weak decay vertex, 
although no quantitative conclusions 
have been drawn, with the exception of \cite{Politzer:1991au}. 
Recently, a systematic 
investigation of the heavy quark limit has been undertaken 
\cite{Beneke:1999br}. 
The conclusion is that all soft and collinear configurations can be 
absorbed either into light cone distribution amplitudes or a form 
factor for the transition $B\to M_1$, where $M_1$ is the meson 
that picks up the light spectator quark. In particular, the corrections 
conventionally termed ``non-factorizable'' are dominated by hard gluon 
exchange and hence computable. The proposed factorization theorem,  
applicable to heavy-light final states ($D\pi$, etc.) and light-light final 
states ($\pi\pi$, $\pi K$, etc.), reads
\begin{eqnarray}
\label{fff}
{\cal A}(B\to M_1 M_2) &=& 
F_{B\to M_1}(0)\,\int\limits_0^1 \!dx\,T^I(x)\,\Phi_{M_2}\!(x) 
\nonumber
\\[-0.3cm]
&&\hspace*{-1.5cm}
+\,\int\limits_0^1 \!d\xi dx dy \,T^{II}(\xi,x,y)\,
\Phi_B(\xi)\Phi_{M_1}\!(y)\Phi_{M_2}\!(x),
\end{eqnarray} 
with corrections that are suppressed as $\Lambda_{\rm QCD}/m_b$. The second 
term is present only for light-light final states and has the form 
of a standard BL-type term. It accounts for hard gluon interactions with 
the spectator quark \cite{Szczepaniak:1990dt}.

The implications of (\ref{fff}) taken at face value are far-reaching. 
Since non-per\-tur\-ba\-tive form factors and light cone distribution 
amplitudes can either be measured or determined in principle 
with lattice QCD, 
the strong phases $\delta_{1,2}$ and amplitudes $A_{1,2}$ are completely 
predicted. CKM parameters can then be directly extracted from 
measurements of branching fractions and CP asymmetries.

Some work remains to be done to demonstrate that (\ref{fff}) gives 
accurate predictions at the $b$ quark scale. A factorization proof to all 
orders has yet to be given, which may imply that an integration over 
intrinsic transverse momentum in the $B$ meson has to be added to 
the second term of (\ref{fff}). Power corrections  in 
$\Lambda_{\rm QCD}/m_b$ can turn out uncomfortably large, if enhanced 
by small current quark masses. It is also worth noting that the 
light cone properties of $B$ mesons have remained largely unexplored. 
In any event, the new approach improves over naive factorization, 
which has been the most commonly used theoretical tool. Because of its 
potential for $B$ factories, applications need to be carefully examined.

\section*{Conclusion}

QCD is a lively field of incredible variety. It is also often technical. 
Comparing today's QCD overviews with the discussion of big ideas 
25 years ago, this variety may even appear intimidating. But this 
transformation in style reflects like no other indicator the progress 
in understanding how QCD works. The challenges provided by strong 
coupling have led to insights 
into how field theory works unparalleled by any other theory. Given the 
intrinsic beauty and simplicity of QCD, together with its role in the 
future high energy physics programme at the energy frontier, we 
can be sure of further progress in the field. \\[0.1cm]


\noindent
I could not have given this talk without the help of many collegues. 
I benefitted in particular from discussions with G.~Buchalla, 
S.~Catani, J.~Forshaw, A.~Hebecker, M.~Kr\"amer, Z.~Kunzst, G.~Salam 
and T.~Teubner.



\def\Discussion{
\setlength{\parskip}{0.3cm}\setlength{\parindent}{0.0cm}
     \bigskip\bigskip      {\Large {\bf Discussion}} \bigskip}
\def\speaker#1{{\bf #1:}\ }

\Discussion

\speaker{G\"unter Grindhammer (MPI, Munich)}
Considering deep inelastic scattering at low $x$, what happens with the
difference between the  BFKL and DGLAP approaches
in the case of heavy quark production?  In this case one has two hard scales,
$Q^2$ and the quark mass.\\

\noindent The answer depends on where the heavy quarks are. If the heavy 
quarks couple to the virtual photon (``the top of the ladder''), there 
is no difference to inclusive DIS at small $x$. If the heavy quark pair 
is coupled to the bottom of the ladder, both ends are perturbative, as 
is the case for forward jet production at $p_T\gg \Lambda_{\rm QCD}$. 
In this case there is no reason to believe that DGLAP evolution should 
be relevant.

\end{document}